\documentclass[onecolumn]{aastex61}
\usepackage{amsmath}

\usepackage{amssymb,fancyhdr,graphicx}

\shorttitle{Toward a Measurement of the Transverse Peculiar Velocity of Galaxy Pairs}
\shortauthors{Truebenbach \& Darling}

\begin{document}
\title{Toward a Measurement of the Transverse Peculiar Velocity of Galaxy Pairs}

\correspondingauthor{Alexandra E. Truebenbach}
\email{alexandra.truebenbach@colorado.edu}

\author{Alexandra E. Truebenbach}
\affil{Center for Astrophysics and Space Astronomy, \\
		Department of Astrophysical and Planetary Sciences, \\
		University of Colorado, 389 UCB, Boulder, CO 80309-0389, USA}

\author{Jeremy Darling}
\affil{Center for Astrophysics and Space Astronomy, \\
	Department of Astrophysical and Planetary Sciences, \\
	University of Colorado, 389 UCB, Boulder, CO 80309-0389, USA}

\begin{abstract}
The transverse peculiar velocities caused by the mass distribution of large-scale structure provide a test of the theoretical matter power spectrum and the cosmological parameters that contribute to its shape. Typically, the matter density distribution of the nearby Universe is measured through redshift or line-of-sight peculiar velocity surveys. However, both methods require model-dependent distance measures to place the galaxies or to differentiate peculiar velocity from the Hubble expansion. In this paper, we use the correlated proper motions of galaxy pairs from the VLBA Extragalactic Proper Motion Catalog to place limits on the transverse peculiar velocity of galaxy pairs with comoving separations $<1500$ Mpc without a reliance on precise distance measurements. The relative proper motions of galaxy pairs across the line of sight can be directly translated into relative peculiar velocities because no proper motion will occur in a homogeneous expansion. We place a $3\sigma$ limit on the relative proper motion of pairs with comoving separations $< 100$ Mpc of $-17.4 \ \mu$as yr$^{-1} < \dot{\theta} / \sin \theta < 19.8 \ \mu$as yr$^{-1}$. We also confirm that large-separation objects ($> 200$ Mpc) are consistent with pure Hubble expansion to within $\sim 5.3 \ \mu$as yr$^{-1}$ ($1 \sigma$). Finally, we predict that {\it Gaia} end-of-mission proper motions will be able to significantly detect the mass distribution of large-scale structure on length scales $< 25$ Mpc. This future detection will allow a test of the shape of the theoretical mass power spectrum without a reliance on precise distance measurements.
\end{abstract}


\section{Introduction}\label{intro}

The modern standard model of cosmology is a spatially flat, expanding Universe dominated by cold dark matter and a cosmological constant. Small matter density perturbations in the primordial Universe drive the initial gravitational collapse of matter and dark matter into overdensities, which give rise to the formation of large-scale structure (LSS), clusters, and galaxies \citep[e.g.,][]{Blumenthaletal1984} through a hierarchical process. The overall process and cosmological parameters that govern the process are, in general, well constrained \citep[e.g.,][]{HuWhite1996a,HuWhite1996b,Huetal1997,Riessetal2001,Peirisetal2003,Moodleyetal2004,PlanckI2016}. However, the exact values of the parameters remain somewhat uncertain, including the baryon and neutrino density \citep[e.g.,][]{EisensteinHu1999, Eisensteinetal2005}, the Hubble constant \citep[e.g.,][]{Eisensteinetal2005,Riessetal2011,PlanckXIII2016,Riessetal2016,Zhangetal2017,Riessetal2018}, the tensor-to-scalar ratio of primordial gravitational waves \citep[e.g.,][]{PlanckXX2016}, and the spatial curvature \citep[e.g.,][]{Eisensteinetal2005,PlanckXIV2016}.

The current distribution of LSS provides a means to test many of the less certain cosmological parameters by comparing the observed LSS to that predicted by cosmological simulations. Most commonly, maps of LSS are created using the sky distribution of visible galaxies with redshift as a proxy for distance \citep[e.g.,][]{deLapparentetal1986,Yorketal2000,Gottetal2005}. However, these maps rely on the assumption that light and visible matter trace the overall dark matter distribution. In general, it is reasonable to assume that galaxies form at the peaks of dark matter \citep{Kaiser1984}, but a bias model \citep[e.g.,][]{Bardeenetal1986,Coles1993,Fry1996,TegmarkPeebles1998} is still required to translate the observed LSS to the dark matter distributions generated by cosmological simulations. 

Galaxy line-of-sight peculiar velocities are an alternate means to track the dark matter distribution that does not require a translation between light and total mass because peculiar velocities of galaxies are directly caused by the matter density distribution. Line-of-sight peculiar velocities are obtained from the difference between the redshift and a redshift-independent distance. Velocity surveys can probe more distant structures than redshift surveys because objects' velocities can be influenced by distant objects that are outside the range of the survey \citep[e.g.,][]{Doumleretal2013,Tullyetal2014}. The drawback of line-of-sight velocity surveys to detect the dark matter distribution is that small uncertainties in a galaxy's distance can translate to large peculiar velocity uncertainties. Therefore, large samples of galaxy redshifts and distances are needed to statistically detect an average matter density distribution \citep{Tullyetal2014}. 

Both methods of mapping LSS produce similar matter density distributions \citep[e.g.,][]{Straussetal1992,Dekeletal1993,Kitauraetal2012,Courtoisetal2012}, which is a good confirmation of the standard model of LSS evolution through hierarchical growth and gravitational instability. However, both methods require redshifts or other model-dependent distance measures to either spatially place the galaxies or to translate spectroscopic line shifts into peculiar velocities. Therefore, another method that is independent of the ``distance ladder'' and other distance models is needed to provide an independent test of the model of LSS evolution.  

Extragalactic proper motions can be used to test models of LSS evolution without a reliance on the ``distance ladder.'' Like line-of-sight peculiar velocities, transverse peculiar velocities directly probe the matter density distribution and do not require any assumptions about the relative abundances of visible matter and dark matter (or how light traces mass). 
As the Universe expands, both line-of-sight and peculiar velocities will contain peculiar motion from gravitational interactions. 
However, line-of-sight velocities require an independent distance measure to differentiate between Hubble expansion and peculiar velocity. On the other hand, orthogonal velocities across the line of sight (observed as proper motions) are separable from the Hubble expansion because no proper motion will occur in a homogeneous expansion \citep{Nusseretal2012,Darling2013}. 

In this paper, we use the relative proper motion of pairs of extragalactic objects from our VLBA Extragalactic Proper Motion Catalog \citep{TruebenbachDarling2017} to constrain transverse peculiar velocities induced by the mass distribution of LSS. We describe the expected signal in Section \ref{pairwise_theory} and calculate the signal for pairs with separations $< 1500$ Mpc is Section \ref{pairwise_measurement}. In Section \ref{limit_section}, we compare our measurement to that predicted by a transverse peculiar velocity two-point correlation statistic calculated from a $z=0$ matter power spectrum \citep{Darling2018}. Finally, we predict {\it Gaia}'s ability to measure the relative convergence of galaxies on small scales (Sec. \ref{future_LSS}) and discuss future improvements to our catalog (Sec. \ref{conclusions}).  In this paper we assume $H_0 = 70$ km s$^{-1}$ Mpc$^{-1}$ and a flat cosmology with $\Omega_\Lambda = 0.73$ and $\Omega_M = 0.27$.

\section{Pairwise Proper Motion - Theory}\label{pairwise_theory}

Extragalactic proper motions are a combination of intrinsic apparent proper motions that are specific to each individual source and uncorrelated between objects (e.g., radio jets), correlated proper motions from cosmological effects \citep[e.g.,][]{Gwinnetal1997, Quercellinietal2009, Nusseretal2012, Darling2013}, and apparent proper motions from observer-induced signatures such as the secular aberration drift \citep[e.g.,][]{Fanselow1983,Bastian1995,Eubanksetal1995,Soversetal1998,Mignard2002,Kovalevsky2003,KopeikinMakarov2006,Titovetal2011, TitovLambert2013,Xuetal2012,Xuetal2013}. Extragalactic proper motions measured with very long baseline interferometry (VLBI) at radio frequencies typically have uncertainties of tens of $\mu$as yr$^{-1}$ \citep[e.g.,][]{TitovLambert2013,TruebenbachDarling2017} that are much larger than the predicted cosmological proper motions ($\lessapprox 15 \ \mu$as yr$^{-1}$). Therefore, studies using extragalactic proper motions to measure correlated cosmological and observer-induced effects are challenged by randomly-oriented intrinsic apparent proper motions and proper motion uncertainties.

The majority of extragalactic proper motions are measured using VLBI observations of radio-loud quasars.
However, in addition to proper motion caused by
cosmological effects, quasars also have intrinsic proper motions,
predominantly due to the motion of plasma in relativistic jets produced by
the quasars \citep[e.g.,][]{BridlePerley1984,Feyetal1997}. These intrinsic proper motions are random in orientation on the sky. 
To separate intrinsic proper motions from cosmological proper motions, we use the relative proper motions of ``pairs'' of quasars \citep{Darling2013}. The intrinsic proper motions of quasars are uncorrelated, whereas LSS growth will display a {\it correlated} signal
in close-separation pairs along the pair axis. With a large enough sample size and a long enough observing period, the contaminating signal caused by intrinsic motions can be reduced enough to detect the correlated proper motion caused by the mass distribution of LSS.

For a ``pair'' of extragalactic objects\footnote{``Pair'' signifies any two randomly selected objects and does not denote any association between the two objects.}, the angular separation of the objects is defined as
\begin{equation}\label{theta}
\sin \theta = \frac{l}{D_A},
\end{equation}
where $l$ is the proper length separating the objects and $D_A$ is the angular diameter distance. Then, the relative proper motion of the two objects along a great circle, $\dot{\theta}$, is
\begin{equation}\label{thetadot}
\frac{\dot{\theta}}{\sin \theta} = \frac{-\dot{D_A}}{D_A} + \frac{\dot{l}}{l} = \frac{-H(z)}{1+z} + \frac{\dot{l}}{l},
\end{equation}
where 
\begin{equation}
H(z) = H_0 \sqrt{\Omega_{M,0}(1+z)^3 + \Omega_\Lambda},
\end{equation}
and $\dot{l}$ is the change in proper length \citep{Darling2013}. See \citet{Darling2013} for further discussion.

If the pair of extragalactic objects are far enough apart that they do not interact gravitationally, then 
\begin{equation}\label{null}
\frac{\dot{l}}{l} = \frac{H(z)}{1+z},
\end{equation}
because the space between the objects expands with the Hubble expansion. Combining this equation and Equation \ref{thetadot} gives $\dot{\theta} = 0$ -- no proper motion is expected for objects comoving with a homogeneously expanding Universe. On the other hand, if the pair are part of a static structure (e.g., two galaxies in the same virialized cluster) then $\dot{l} / l = 0$ and the pair will appear to converge as they move away from us with the Hubble flow. For a static structure at $z=0$, the apparent convergence of a pair can be approximated as $H_0 \sim 15 \ \mu$as yr$^{-1}$. The majority of extragalactic pairs have large physical separations and will show the null signal expected for gravitationally non-interacting objects entrained in the Hubble flow. However, pairs with close physical separations ($\lesssim 50$ Mpc, comoving) are expected to show relative proper motion that is a combination of the two signals; the pair will both appear to converge as the galaxies move away from us with the Hubble flow and they will have relative proper motion determined by their gravitational interaction with their local mass density distribution. Therefore, most pairs are expected to have a total relative proper motion that is less than the $\sim 15 \ \mu$as yr$^{-1}$ signal predicted for static structures.

\section{Pairwise Proper Motion - Measurement}\label{pairwise_measurement}
We measured the relative proper motion of extragalactic objects using the VLBA Extragalactic Proper Motion Catalog \citep{TruebenbachDarling2017}. The catalog contains 713 proper motions measured with VLBI, 596 of which have well measured redshifts and are therefore usable for this measurement. Redshifts are required to calculate physical separations between pairs. We excluded all catalog objects with either no redshift or a flagged redshift in the VLBA Extragalactic Proper Motion Catalog. A flagged redshift indicates that the redshift is either photometric or uncertain.

Prior to calculating pairwise proper motions, we also fit and subtracted the secular aberration drift signal measured in our catalog \citep{TruebenbachDarling2017}. The secular aberration drift, an apparent curl-free dipole motion of all extragalactic objects caused by the Solar System's acceleration towards the Galactic center, is an observer-induced apparent extragalactic motion and must be removed before analyzing any proper motions caused by cosmological effects. In \citet{TruebenbachDarling2017}, we measured a dipole signal from the aberration drift with a square root power equal to $4.89 \pm 0.77 \ \mu$as yr$^{-1}$ and an apex of ($275.2 \pm 10.0^\circ$, $-29.4 \pm 8.8^\circ$) in equatorial coordinates (see \citet{Xuetal2013} and \citet{TitovLambert2013} for other measurements of the secular aberration drift). Although we only detect a portion of the predicted dipole signal \citep[expected square root power $= 15.6 \pm 2.3 \ \mu$as yr$^{-1}$; derived from][]{Reidetal2009} in our catalog, we still must remove this partial dipole before proceeding. 

For all objects, we calculated their relative proper motion, $\dot{\theta}/ \sin \theta$, with respect to all other sources in the catalog. For two points on a sphere, the angular separation of the two points is 
\begin{equation}\label{separation}
\theta_{ij} = \arccos ( \sin \delta_i \ \sin \delta_j + \cos \delta_i \ \cos \delta_j \ \cos[\alpha_i - \alpha_j]),
\end{equation}
where the points' equatorial coordinates are ($\alpha_i$,$\delta_i$). The change in the angular separation due to the proper motion of those points is
\begin{multline}\label{thetadot_eqn}
\dot{\theta}_{ij} = -(\cos \delta_i \ \sin \delta_j [\mu_{\delta,i} - \mu_{\delta,j} \cos(\alpha_i - \alpha_j)] + \sin \delta_i \ \cos \delta_j [\mu_{\delta,j} - \mu_{\delta,i} \cos (\alpha_i - \alpha_j)] \\
- \cos \delta_i \ \ cos \delta_j \ \sin(\alpha_i - \alpha_j)[\mu_{\alpha,i} - \mu_{\alpha,j}])/ \sin \theta_{ij}
\end{multline} 
\citep{Darling2013}.

Comoving separation is calculated from the angular separation and the cosine rule:
\begin{equation}
D_{ij}^2 = D_i^2 + D_j^2 - 2 D_i D_j \cos \theta_{ij}.
\end{equation}
$D_i$ and $D_j$ are the comoving proper distances of the objects, calculated by integrating the inverse of the scale factor from the object's time of emission to the present:
\begin{equation}
D_i =c \int_{t_i}^{0} \frac{dt'}{a(t')}.
\end{equation}

Several galaxies (M81, M84, 3C274, 1923+210, and NGC4261) are too nearby for redshift to be a good proxy for distance. For these, we used a median of various redshift-independent distances reported by the NASA/IPAC Extragalactic Database\footnote{https://ned.ipac.caltech.edu} (NED), including distances measured from Type 1a supernovae light curves \citep[e.g.,][]{Colgate1979, Riessetal1996, Perlmutteretal1999}, globular cluster luminosity functions \citep[e.g.,][]{Hanes1977, HarrisRacine1979}, surface brightness fluctuations \citep{TonrySchneider1988, Tonryetal1990, Jensenetal1998, Blakesleeetal1999, Meietal2005}, and the Faber-Jackson relation \citep{FaberJackson1976}. We did not include distance measurements made prior to the year 2000.

\begin{figure}
	\label{pairwise_fig}
	\centerline{\includegraphics[width=.7\textwidth]{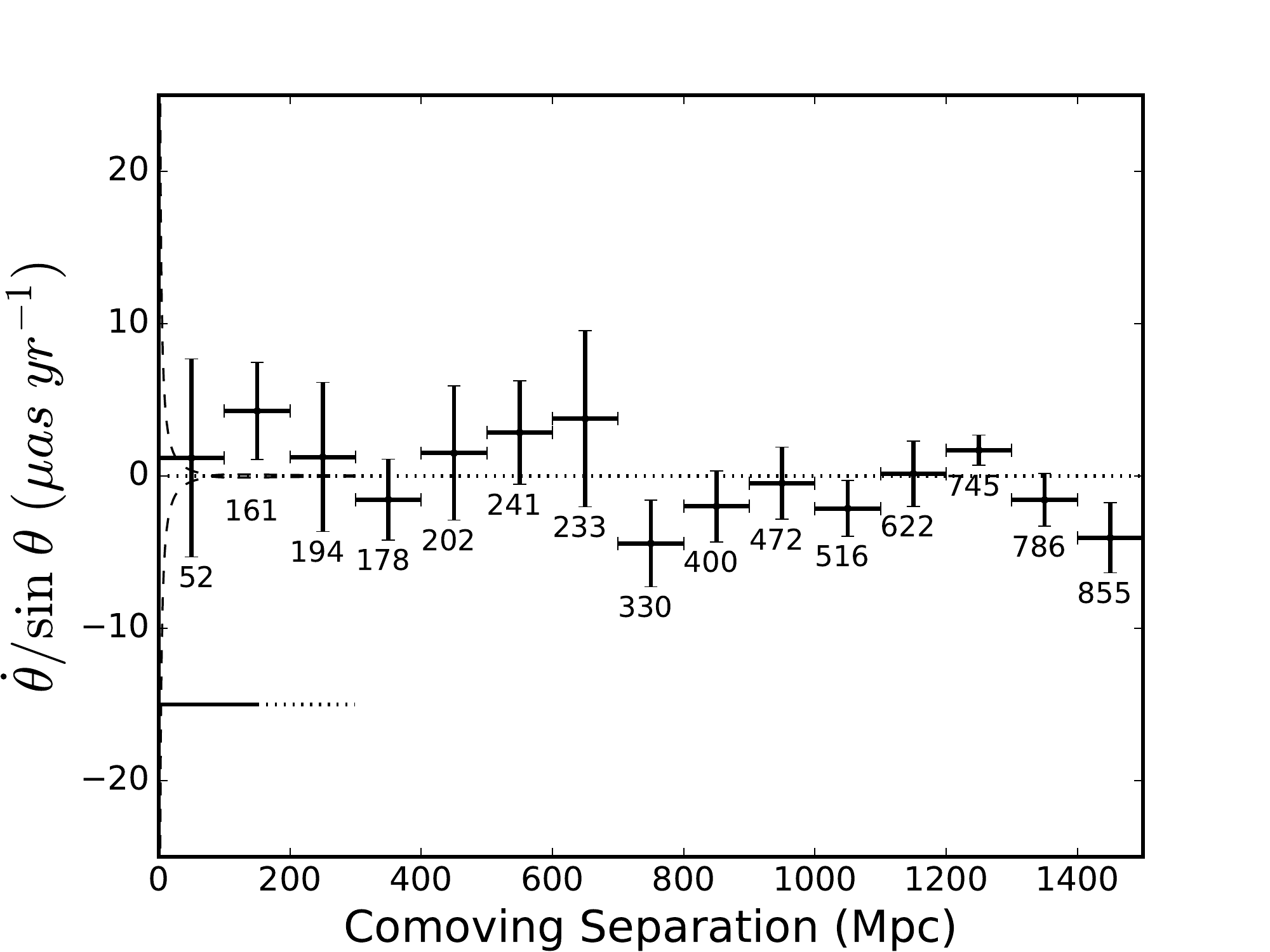}}
	\caption{Measured pairwise divergence or convergence vs. comoving separation of radio sources in the proper motion catalog. Numbers beneath each data point indicate the number of pairs in each bin. The dotted line shows the null signal expected for gravitationally non-interacting pairs expanding with the Hubble flow, while the solid line at $\dot{\theta}/ \sin \theta = -15 \ \mu$as yr$^{-1}$ indicates the expected convergence of a static structure at $z=0$. The dashed curved lines show the predicted relative proper motion of a pair with an angular separation of 30$^\circ$ derived from the transverse, peculiar velocity two-point correlation function \citep[Eqn \ref{xi_vdot} and][]{Darling2018}. We find a 3 sigma limit on the convergence rate of gravitationally-interacting objects ($< 100$ Mpc) of $-17.4 \ \mu$as yr$^{-1} < \dot{\theta} / \sin \theta < 19.8 \ \mu$as yr$^{-1}$. Large-separation pairs are consistent with the null signal expected for Hubble expansion to within $1\sigma$: $< 2.3 \ \mu$as yr$^{-1}$ for pairs with separations $>800$ Mpc and $< 5.7 \ \mu$as yr$^{-1}$ for pairs with separations $200-800$ Mpc.}
\end{figure}

Figure \ref{pairwise_fig} shows the relative proper motion of radio source pairs binned by comoving separation. The relative proper motions and error bars are the median and standard deviation of a bootstrap distribution (see below). The solid line at $\dot{\theta}/ \sin \theta = -15 \ \mu$as yr$^{-1}$ indicates the expected convergence of a static structure at $z=0$. This represents the most extreme deviation from the null signal of pure Hubble expansion for pairs whose relative proper motion is not dominated by intrinsic motions (i.e., radio jets). The curved lines show the predicted relative proper motion as a function of pair separation estimated as the square root of the transverse peculiar velocity two-point correlation function for a pair with an angular separation of 30 degrees \citep[see Section \ref{limit_section} and][]{Darling2018}. It is important to note that individual objects can be part of many pairs, therefore the bins in Figure \ref{pairwise_fig} are not statistically independent.

Table \ref{pairwise_table} lists the bin medians and uncertainties. As predicted, bins with comoving separations $> 100$ Mpc show no relative motion and are consistent with pure Hubble expansion. Despite the large scatter of relative proper motions within each bin due to the intrinsic proper motions from jets, we are still able to measure the expected cosmological behavior of large-separation pairs to high precision. Large-separation pairs are consistent with pure Hubble expansion to within $1\sigma$: $< 2.3 \ \mu$as yr$^{-1}$ for pairs with separations $>800$ Mpc and $< 5.7 \ \mu$as yr$^{-1}$ for pairs with separations $200-800$ Mpc. This result is a strong verification of our ability to place limits on small relative proper motion signals from a sample of large, uncorrelated intrinsic proper motions.  

\begin{deluxetable}{llrl}
	\tablecaption{Binned Pairwise Proper Motions}
	\tablewidth{0pc}
	\tablehead{\colhead{Comoving} & \colhead{$\langle z \rangle$} & \colhead{$\langle \dot{\theta} / \sin \theta \rangle$} & \colhead{N} \\
		\colhead{Separation} & \colhead{}  & \colhead{}  & \colhead{}   \\
		\colhead{(Mpc)} & \colhead{} & \colhead{($\mu$as yr$^{-1}$)} & \colhead{} }
	\startdata
	0--100 & 0.10 & 1.2 (6.2) & 52 \\
	100--200 & 0.38 & 4.3 (3.0) & 161 \\
	200--300 & 0.21 & 1.3 (5.0) & 194 \\
	300--400 & 0.49 & -1.7 (4.2) & 178 \\
	400--500 & 0.62 & 1.7 (4.2) & 202 \\
	500--600 & 0.57 & 2.6 (3.3) & 241 \\
	600--700 & 0.65 & 3.7 (5.7) & 233 \\
	700--800 & 0.67 & -4.2 (2.7) & 330 \\
	800--900 & 0.77 & -2.0 (2.2) & 400 \\
	900--1000 & 0.77 & -0.9 (2.3) & 472 \\
	1000--1100 & 0.83 & -2.2 (1.7) & 516 \\
	1100--1200 & 0.86 & 0.2 (2.2) & 622 \\
	1200--1300 & 0.89 & 1.7 (0.9) & 745 \\
	1300--1400 & 0.95 & -1.3 (1.7) & 786 \\
	1400--1500 & 0.96 & -4.0 (2.3) & 855 \\
	\enddata
	\label{pairwise_table}
	\tablecomments{Columns from left to right: Pair comoving separation, average redshift of pairs within separation range, median relative proper motion of the bootstrap distribution, and number of pairs within separation range. Parenthetical values indicate 1$\sigma$ standard deviation of the bootstrap distribution.}
\end{deluxetable}

The bin with comoving separations $< 100$ Mpc is consistent with both Hubble expansion and the maximum convergence of static structures at $z=0$. The predicted relative proper motion from the two-point correlation function lies between these two expected signals and is also consistent with the first bin. Overall, the $<100$ Mpc bin contains too few pairs to detect any significant deviation from pure Hubble expansion. Additionally, Figure \ref{pairwise_fig} shows that the predicted two-point correlation function only deviates from zero for pairs with comoving separations $\lessapprox 50$ Mpc. There are only 8 pairs with separations $< 50$ Mpc, with a mean relative proper motion of $-11.5  \ \mu$as yr$^{-1}$ and a standard deviation of $22.2 \ \mu$as yr$^{-1}$. Therefore, there are too few pairs with separations $<50$ Mpc to statistically differentiate between pure Hubble expansion and the convergence expected for small-separation pairs.

\begin{figure}
	\centerline{\includegraphics[width=0.6\textwidth, trim= 3cm 4.5cm 1.5cm 3cm, clip]{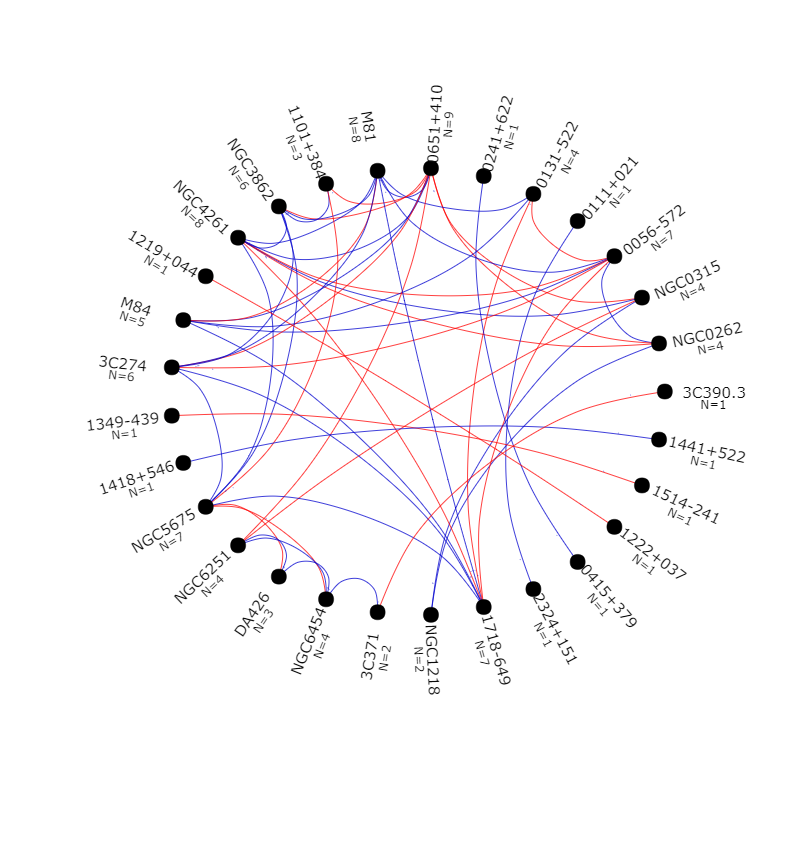}}
	\caption{Chord diagram showing all objects that have pairs within 100 Mpc, comoving. The lines connect all objects that are pairs with each other. Blue lines indicate converging pairs ($\dot{\theta} / \theta < 0$), while red lines indicate diverging pairs ($\dot{\theta} / \theta > 0$). The number of pairs are also listed underneath each object. All pairs whose relative velocity is three times greater than the expected relative velocity of a static structure entrained in the Hubble flow (Eqn. \ref{vmax}) are removed from the plot.
		\label{chord}}
\end{figure}

The primary concern when calculating binned pairwise proper motion is the effect of individual objects with well-measured, large, intrinsic proper motions from relativistic jets. These intrinsic proper motions are often highly significant outliers in the bin and can dominate an error-weighted least-squares fit. This effect is particularly a concern for the bin of pairs with comoving separations $<100$ Mpc. Figure \ref{chord} shows the objects that comprise the first bin along the outer circle, while the connecting chords show the pairs. This figure demonstrates that many of the pairs in the first bin are composed of the same small number of galaxies. If one galaxy that is in many pairs has a significant intrinsic proper motion, it can have a large effect on the bin's average.

\begin{figure}
	\centerline{\includegraphics[width=.6\textwidth]{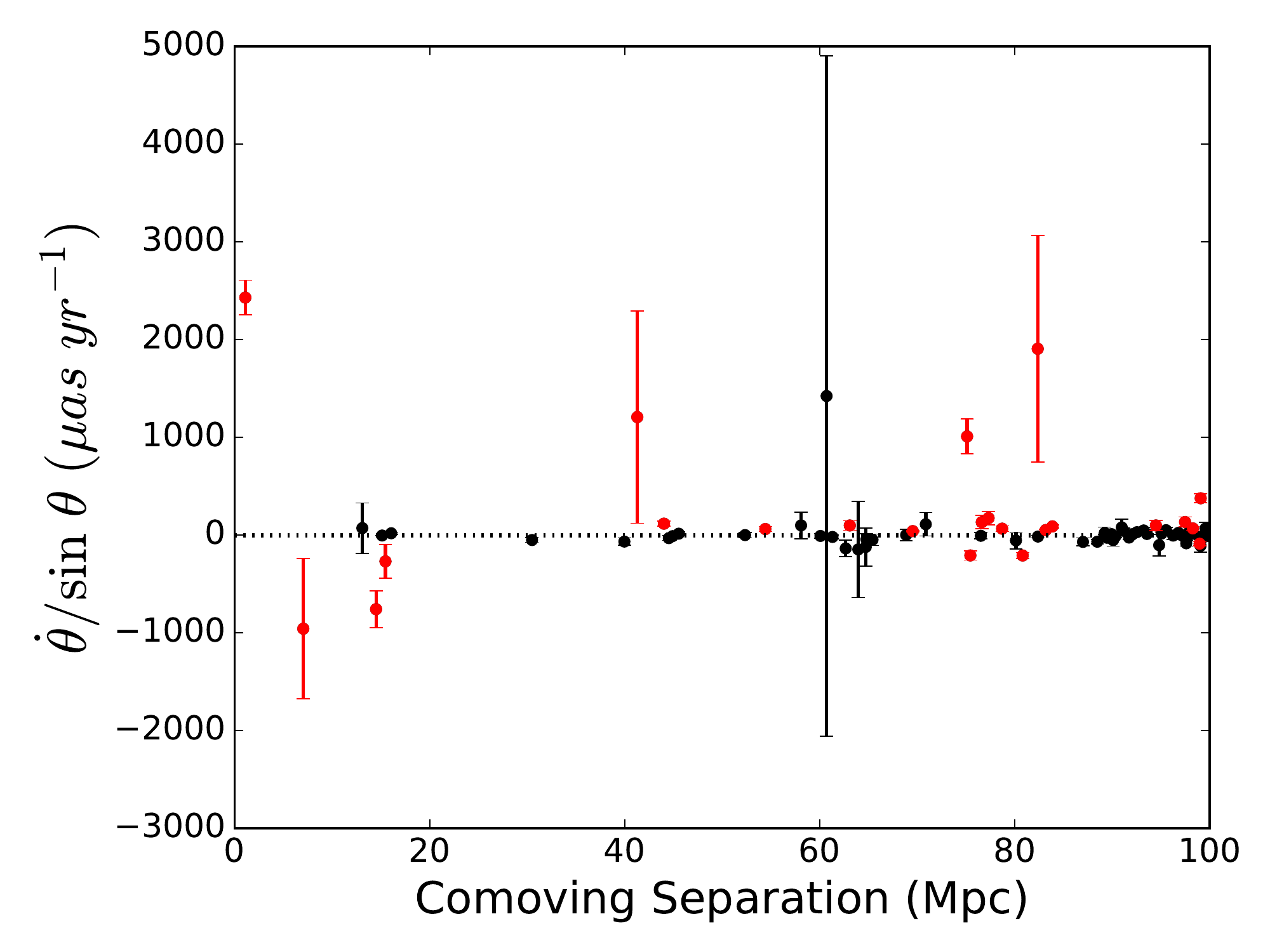}}
	\caption{Measured pairwise divergence or convergence vs. comoving separation for all individual pairs with comoving separations $<100$ Mpc. Black points are pairs included in the bin median, while red points are pairs that were clipped because their relative velocity is more than 1$\sigma$ greater than three times the expected relative velocity for a static structure at the pair redshift. The dashed line shows the null signal expected for large-separation, gravitationally non-interacting pairs expanding with the Hubble flow.
		\label{pairwise-individual}}
\end{figure}

We use relative proper motions along a pair axis to somewhat mitigate this effect. Intrinsic proper motions are randomly oriented and are uncorrelated between objects, reducing the likelihood that the intrinsic proper motion will lie along the pair axis. However, Figure \ref{pairwise-individual} shows the individual pairwise proper motions for all pairs with comoving separations $<100$ Mpc and illustrates that the bin uncertainty is still dominated by significant individual pairwise proper motions. To quantify this scatter and reduce the influence of individual galaxies on the bin average, the  bin averages and uncertainties in Figure \ref{pairwise_fig} are the median and standard deviation of 1,000 iterations of a bootstrap distribution where subsets of the galaxies that comprise the bin are selected. In addition, the bin uncertainties must be bootstrapped because the bins are correlated -- multiple bins contain many of the same objects. Therefore, we cannot directly calculate an uncertainty for each bin without also accounting for its correlation to the other bins. In this case, a bootstrap distribution is an accurate way to estimate the bin uncertainties.

 We also reduce the influence of significant outliers by calculating the bin average of each bootstrap binning iteration through a maximum-likelihood ``permissive fit'' method \citep[see][]{Darlingetal2017}. This method allows for highly significant outlier data points by assuming that the offset between model and data will in some cases be bounded by the measured uncertainty \cite[p. 168]{SiviaSkilling2006}. For each iteration, we use a least-squares technique with {\tt lmfit} \citep{Newvilleetal2014} to obtain the bin average that minimizes the permissive fit residual.

Before we bootstrap and fit each bin, we clip all pairs whose relative velocity is greater than three times that expected for a static structure at the pair's redshift. For a pair at an average comoving proper distance $D$, with relative apparent velocity $v = (\dot{\theta} / \sin \theta) \ D$, the maximum convergence expected for a static pair that is entrained in the Hubble flow is 
\begin{equation}
\frac{\dot{\theta}}{\sin \theta} = \frac{- H(z)}{1+z},
\end{equation}
which equates to a maximum expected relative velocity of 
\begin{equation}\label{vmax}
v_{\rm{max}} =  \frac{- H(z)}{1+z} \ D.
\end{equation}
We include in our bin averages all pairs whose relative apparent velocity is within $1\sigma$ of the range $3 v_{\rm{max}} < v < -3 v_{\rm{max}}$, thereby removing all pairs that are contracting / expanding more than three times faster than the contraction expected for a static structure receding in the Hubble flow. This clipping of large apparent relative velocities removes pairs whose relative proper motions are jet-dominated rather than dominated by peculiar velocities caused by the mass distribution of LSS. Figure \ref{pairwise-individual} shows the clipped pairs from the first bin in red. This clipping criterion is preferable over simpler criteria (e.g., clipping objects with absolute proper motions or proper motion uncertainties $> 100 \ \mu$as yr$^{-1}$) because it is distance independent. If we clip objects with large absolute proper motions and proper motion uncertainties, we remove the most distant objects from our bin averages because peculiar velocity uncertainty is higher for more distant objects; these criteria bias the bin averages towards nearby pairs. On the other hand, our criterion clips using a $1\sigma$ deviation from the maximum expected relative velocity determined from the Hubble flow, which scales with distance. 

Finally, we include three rounds of $5\sigma$ clipping of pairs that deviate from the bin median. On average, this iterative clipping removes $< 1$ pair from each bootstrap iteration for bins with separations $< 500$ Mpc. It removes $1 - 6$ pairs for bins with separations $> 500$ Mpc.

\section{Limit on Peculiar Transverse Velocity}\label{limit_section}
As discussed in Section \ref{pairwise_measurement} and shown in Figure \ref{pairwise_fig}, the bin of pairs with comoving separations $<100$ Mpc is consistent with both pure Hubble expansion and the expected convergence of a static structure at $z=0$. It is also consistent with the predicted relative proper motion of a pair with an angular separation of $30^\circ$ derived from the transverse peculiar velocity two-point correlation function (Eqn \ref{xi_vdot}). Based on the $<100$ Mpc bin median, we find a $3 \sigma$ constraint on the convergence rate of gravitationally-interacting objects of $-17.4 \ \mu$as yr$^{-1} < \dot{\theta} / \sin \theta < 19.8 \ \mu$as yr$^{-1}$. Large-separation pairs are consistent with pure Hubble expansion to within $1\sigma$: $< 2.3 \ \mu$as yr$^{-1}$ for pairs with separations $>800$ Mpc and $< 5.7 \ \mu$as yr$^{-1}$ for pairs with separations $200-800$ Mpc.

The curved line representing the expected signal from the correlation function in Figure \ref{pairwise_fig} is a rough approximation meant to aid in the reader's interpretation of the plot. The expected pair proper motion for close-separation pairs can be more precisely calculated using a two-point correlation statistic and a theoretical mass power spectrum. For a pair of objects with transverse peculiar velocities $\vec{v}_{\bot,1}$ and $\vec{v}_{\bot,2}$, respectively, \citet{Darling2018} define a two-point correlation statistic,
\begin{equation}\label{xi_vdot}
\xi_{v,\bot}(\vec{D}_1,\vec{D}_2) = \langle (\vec{v}_{\bot,1} \cdot \hat{x})  (\vec{v}_{\bot,2} \cdot \hat{x}) \rangle,
\end{equation}
where the angle brackets represent summing over all pairs with a given separation, $\vec{D}_1$ and $\vec{D}_2$ are the radial vectors to the objects, and $\hat{x}$ is the unit space vector connecting the two radial vectors. In Appendix \ref{appendix}, we calculate $\xi_i$, the transverse peculiar velocity two-point correlation statistic for each individual pair.

\begin{figure}
	\centerline{\includegraphics[width=.7\textwidth]{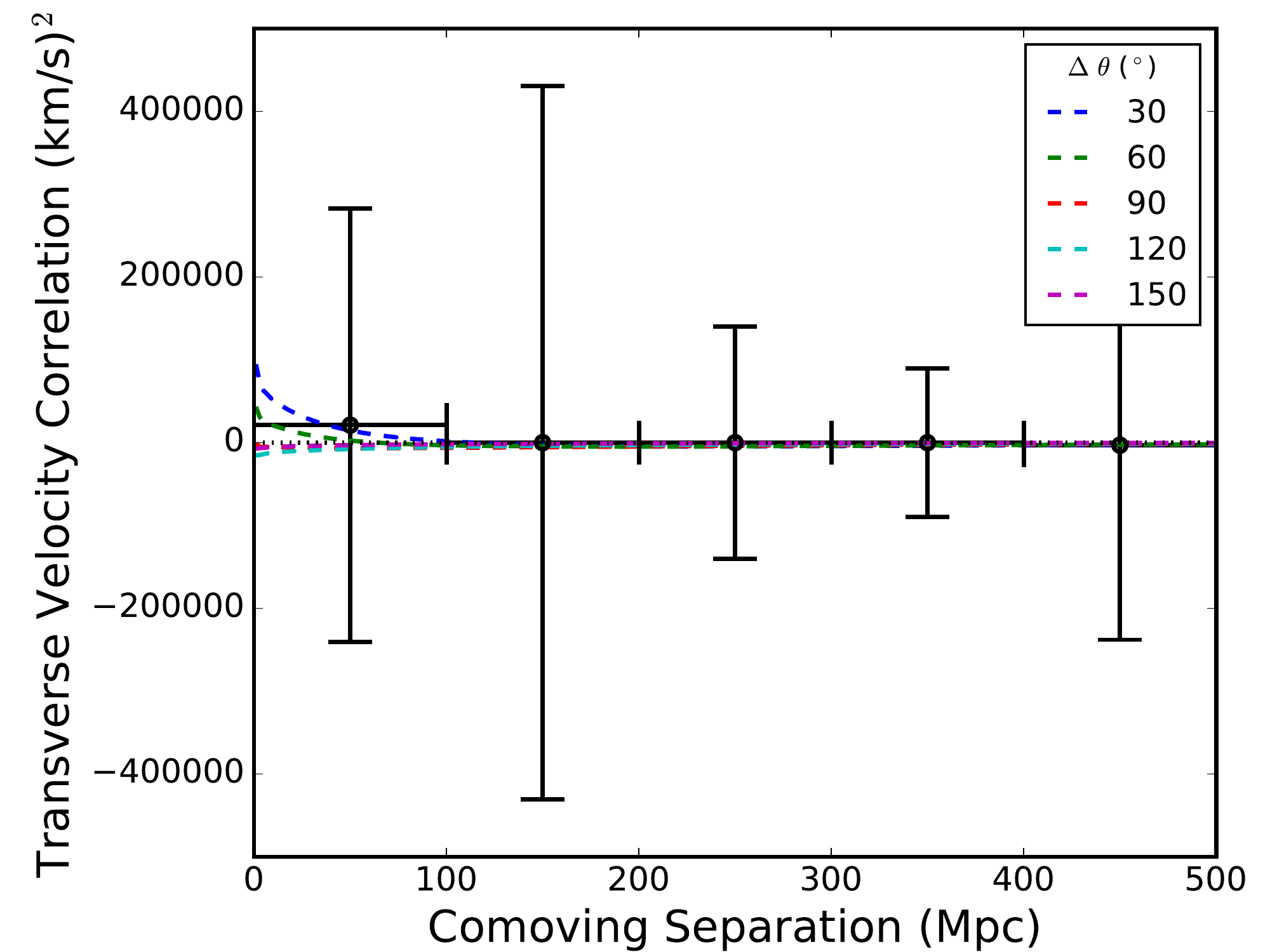}}
	\caption{The average individual transverse peculiar velocity two-point correlation statistic (Eqn \ref{xi_i}) for all pairs of objects with comoving separations $< 500$ Mpc. The expected correlation statistic from the theoretical mass power spectrum is plotted for several angular separations as dashed colored lines. The dotted black line is the expected null signal for pure Hubble expansion. The number of pairs per bin is the same as for Figure \ref{pairwise_fig}.
		\label{xi_plot}}
\end{figure}

Figure \ref{xi_plot} shows the median of the bootstrap distribution for the individual two-point correlation statistic (Eqn \ref{xi_i}) for all pairs of objects with comoving separations $< 1500$ Mpc. The pairs are selected and clipped in the same manner as was used to create Figure \ref{pairwise_fig}. Again, the bin centers and error bars are the median and standard deviation of a 1000 iteration bootstrap distribution. The expected correlation statistic from the theoretical mass power spectrum is also plotted for several angular separations. Like Figure \ref{pairwise_fig}, this figure shows that the bin with comoving separations $< 100$ Mpc is consistent with both pure Hubble expansion and the predicted relative motions of close-separation pairs from the mass power spectrum at several angular separations. The $<100$ Mpc bin contains a roughly even distribution of pair angular separations between $0^\circ$ and $180^\circ$ (most pairs in the bin include very nearby galaxies within 200 comoving Mpc of our galaxy). The plotted correlation statistics show that the expected $\xi_i$ vary widely as a function of angular separation, therefore the bin median is somewhat diluted by a wide range of $\xi_i$ values. In future studies with larger sample sizes, it may be beneficial to bin pairs by angular separation.

\section{Future Limits on Peculiar Transverse Velocity with {\it Gaia}}\label{future_LSS}
Our current catalog of extragalactic proper motions is not sufficiently large to detect the average motion caused by the mass distribution of LSS. Specifically, we lack sufficient pairs with comoving separations $<50$ Mpc, where the predicted correlation of LSS deviates from the null signal predicted on large length scales. If we roughly expect the error bars in Figure \ref{xi_plot} to decrease by $N$, then we would need $\sim 30$ more close-separation pairs to significantly detect the relative proper motion from LSS (this assumes that all new pairs are independent measurements and that multiple pairs do not contain the same object). 

{\it Gaia} \citep{GaiaCollaboration2016, GaiaCollaboration2018} extragalactic proper motions will significantly increase the sample of pairs with separations $<50$ Mpc. However, {\it Gaia} proper motions have much higher uncertainties \citep[$\sim 200 \ \mu$as yr$^{-1}$;][]{Paineetal2017} than those produced through VLBI observations \citep[10's of $\mu$as yr$^{-1}$;][]{TitovLambert2013,TruebenbachDarling2017}. To assess the contribution of {\it Gaia} proper motions to our measurement of the peculiar velocities caused by the mass distribution of LSS, we created a simulated catalog of proper motions from {\it Gaia} DR1 \citep{Lindegrenetal2016}. We cross-matched the DR1 catalog with the third release of the Large Quasar Astrometric Catalog \citep[LQAC3;][]{Souchayetal2009, Souchayetal2015} to select a large sample of extragalactic objects and to obtain redshifts. Using the {\tt pyGaia}\footnote{https://pypi.python.org/pypi/PyGaia} package, we gave each quasar a predicted {\it Gaia} end-of-mission proper motion uncertainty based on its ecliptic angle and optical {\it G}-band magnitude. Then, we gave each quasar a proper motion randomly drawn from the distribution of predicted proper motion uncertainties in order to create a null proper motion noisy data set for signal recovery simulations. Our total catalog roughly follows the Sloan Digital Sky Survey \citep[SDSS;][]{Yorketal2000} footprint with additional sparse coverage across the entire sky and contains 189,562 objects with an average proper motion uncertainty of $\sim 210 \ \mu$as yr$^{-1}$.

We found 236,218 pairs with comoving separations $<50$ Mpc. For each pair with a separation $<100$ Mpc (610,841 pairs), we assumed that the pair was at the average comoving distance of the pair constituents and added the expected correlation statistic for equidistant pairs  predicted from the matter power spectrum to the correlation statistic calculated from the pair's random noise-consistent proper motions. Then we attempted to recover the injected signal to predict whether {\it Gaia} proper motions will be able to detect the transverse peculiar velocities induced by the mass distribution of LSS for close-separation pairs. Note that adding two correlation statistics rather than adding two velocities and then calculating the correlation statistic underestimates the total injected signal. However, it is non-trivial to recover predicted velocities from the injected correlation statistic. Therefore, we use the approximation of added correlation statistics, which is sufficient for this initial estimate. Additionally, the assumption of equidistant pairs overestimates the correlation statistic compared to a full calculation of the statistic for randomly oriented pairs \citep{Darling2018}. But, again, this approximation of sufficient for this initial estimate.

\begin{figure}
	\centerline{\includegraphics[width=.7\textwidth]{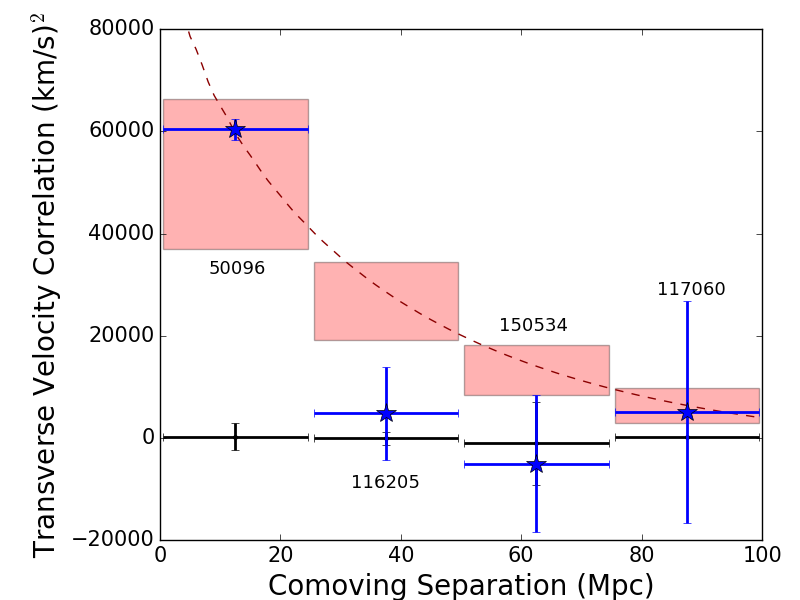}}
	\caption{The transverse peculiar velocity two-point correlation statistic (Eqn \ref{xi_i}) for the {\it Gaia} catalog simulated proper motions. The blue stars show the bin averages for the catalog with the expected correlation statistic added to the no-signal statistic calculated from the random noise-consistent proper motions, while the black dots show the bin averages for the catalog with no signal added. The error bars on these points show the 68\% confidence intervals of the probability distributions of the bin averages. The red boxes show the 1$\sigma$ distribution of expected correlation statistics for all pairs in each bin, while the red dashed line shows the expected correlation statistic from the theoretical mass power spectrum for pairs with angular separations of $5^\circ$. The number of pairs per bin is listed above or below each blue star. 
		\label{gaia_sim}}
\end{figure}

\begin{figure}
	\centerline{\includegraphics[width=.6\textwidth]{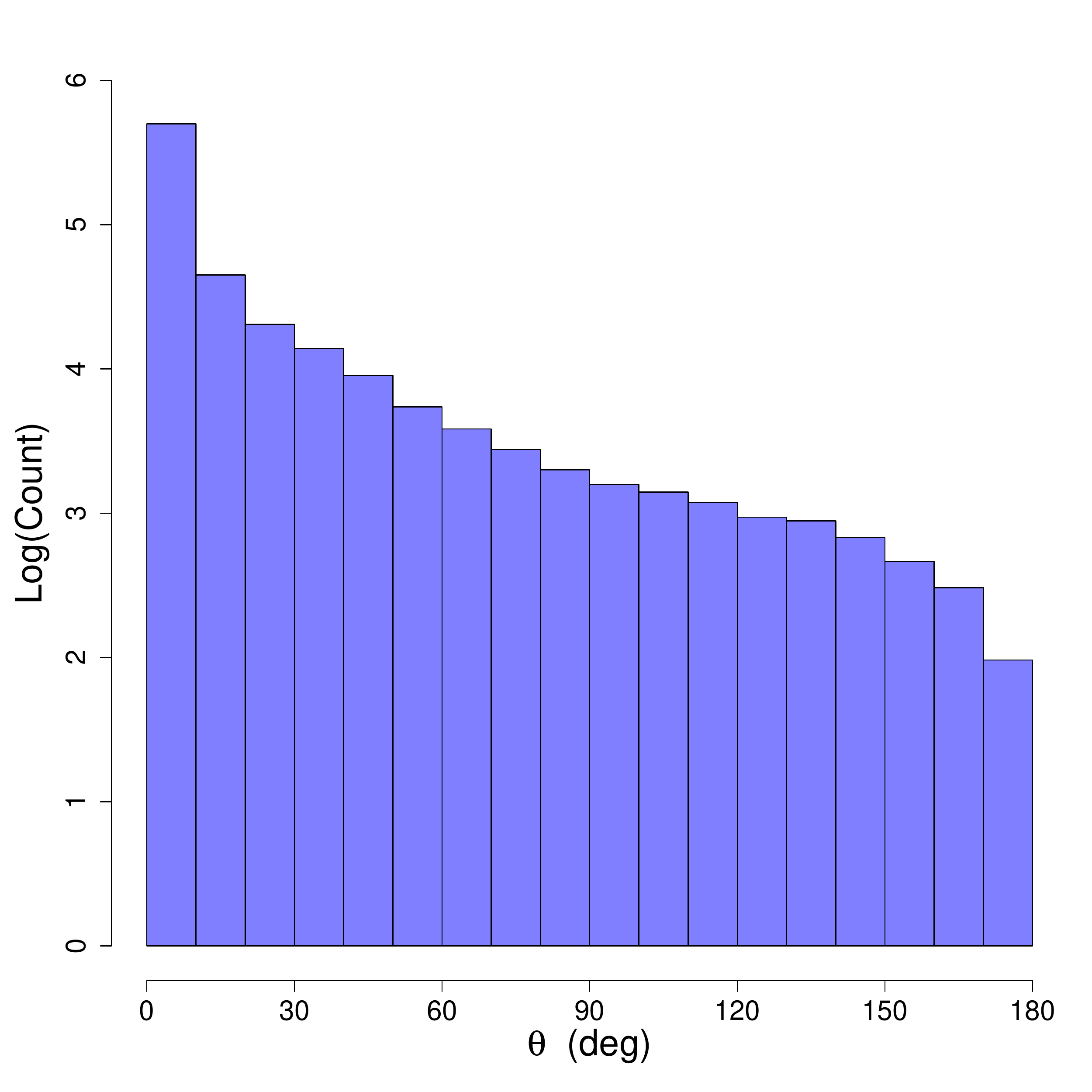}}
	\caption{Histogram showing the distribution of pair angular separations in the {\it Gaia} catalog with simulated proper motions.
		\label{gaia_hist}}
\end{figure}

Figure \ref{gaia_sim} shows the binned correlation statistic for our catalog of simulated {\it Gaia} proper motions. The blue stars show the bin averages for the catalog with the expected correlation statistic added to the no-signal statistic calculated from the random noise-consistent proper motions, while the black dots show the bin averages for the catalog with no signal added. The red boxes show the 1$\sigma$ distribution of expected correlation statistics for all pairs in each bin. We performed the same pairs search and $\xi_i$ binning for both {\it Gaia} catalogs as we did for our proper motion catalog. However, because of the large sample size, we did not include a bootstrap. Instead, the plotted bin values and uncertainties are the maximum likelihood solutions for the bin averages and the 68\% confidence intervals of the probability distributions. Table \ref{gaia_table} lists the bin averages and uncertainties, along with several additional bin statistics, while Figure \ref{gaia_hist} shows the catalog's distribution of pair angular separation. Both the table and figure show that all four bins are dominated by pairs with angular separations less than $10^\circ$. 

The {\it Gaia} catalog simulated proper motions demonstrate that even a subset of 189,562 {\it Gaia} extragalactic proper motions will be able to detect a deviation from pure Hubble expansion for close-separation pairs. The first bin with comoving separations $<25$ Mpc is a 30$\sigma$ detection and is in good agreement with the input signal. Given the approximations made when calculating the correlation statistic, this significance is likely overoptimistic. However, it demonstrates that a significant detection will be possible with {\it Gaia}. Although {\it Gaia} proper motions will be much less precise than VLBI proper motions, the two orders of magnitude increase in the number of available extragalactic proper motions will allow a significant detection of the expected transverse peculiar velocity two-point correlation function at physical separations $< 25$ Mpc. The shape and magnitude of the measured correlation will allow a means to test the shape of the theoretical mass power spectrum that is independent of any distance ladder.

There are several additional sources of proper motion uncertainty for {\it Gaia} extragalactic objects that are not accounted for in our catalog. Intrinsic active galactic nuclei (AGN) variability can cause an apparent proper motion of up to a few mas over the variability timescale for nearby galaxies \citep{Popovicetal2012}. Microlensing can also cause an apparent centroid location shift as large as tens of $\mu$as for stellar mass lensing objects \citep{TreyerWambsganss2004} and a few mas for lensing stellar clusters \citep{PopovicSimic2013}. Both AGN variability and microlensing will add uncorrelated proper motion noise to {\it Gaia} proper motions. However, the expected signal due to the mass distribution of LSS is a correlated, relative proper motion. Therefore, although these effects will add to the overall noise of {\it Gaia} proper motions, they will have a less significant effect on the noise of our measured correlation statistic.

\begin{deluxetable}{crrrrr}
\tablecaption{{\it Gaia} Catalog Simulated Proper Motion Statistics}
\tablehead{\colhead{Comoving Separation} & \colhead{$\langle z \rangle$} & \colhead{Recovered $\langle \xi_i \rangle$} & \colhead{Expected $\langle \xi_i \rangle$} & \colhead{$\langle \theta \rangle$} &  \colhead{N} \\
\colhead{(Mpc)} & \colhead{} & \colhead{(km s$^{-1}$)$^2$} & \colhead{(km s$^{-1}$)$^2$} & \colhead{(deg)} & \colhead{}}
\startdata
0 -- 25 & 0.66 & 6.0 (0.2)  $\times \ 10^4$ & 5.2 (1.5)  $\times \ 10^4$ & 5.2  & 50096 \\
25 -- 50 & 0.72 & 0.5 (0.9) $\times \ 10^4$& 2.6 (0.8)  $\times \ 10^4$ & 7.0 & 116205 \\
50 -- 75 & 0.75 & -0.5 (1.3) $\times \ 10^4$ & 1.3 (0.5)  $\times \ 10^4$ & 7.7 & 150534 \\
75 -- 100 & 0.77 & 0.5 (2.2) $\times \ 10^4$ & 0.6 (0.3)  $\times \ 10^4$ & 9.4 & 117060 \\
\enddata
\label{gaia_table}
\tablecomments{Columns from left to right: Pair comoving separation, average redshift of pairs within separation range, recovered average transverse peculiar velocity two-point correlation statistic and  1$\sigma$ uncertainties in parentheses, expected average correlation statistic based on pair physical and angular separation with the expected distribution standard deviation in parentheses, average pair angular separation, and number of pairs within separation range. }
\end{deluxetable}

\section{Conclusions}\label{conclusions}
In this paper, we measured the relative proper motion of all pairs of objects from the VLBA Extragalactic Proper Motion Catalog \citep{TruebenbachDarling2017} with comoving separations less than 1500 Mpc. We found that large-separation pairs with separations $> 200$ Mpc are consistent with the null signal expected for Hubble expansion to within $\sim 5.7 \ \mu$as yr$^{-1}$ ($1\sigma$). This result demonstrates that our method of measuring the relative proper motion of pairs is able to extract small proper motion signals to high precision despite the large scatter of individual pairs due to the intrinsic proper motions from jets.

We also found that we have too few close-separation pairs ($<100$ Mpc) to statistically detect the expected convergence predicted by the peculiar transverse velocity two-point correlation function. We found a 3$\sigma$ limit on the convergence rate of close-separation pairs of $-17.4 \ \mu$as yr$^{-1} < \dot{\theta} / \sin \theta < 19.8 \ \mu$as yr$^{-1}$. Additionally, we estimated that we would need $\sim 30$ more uncorrelated pairs with separations $<50$ Mpc to make a 3$\sigma$ detection of this effect.  

There are several available avenues for expanding our sample of close-separation pairs. The best source for additional extragalactic proper motions is very long baseline interferometry because of its uniquely high angular resolution. From the uncertainties in Figure \ref{xi_plot}, we predict that we would need $\sim 30$ additional pairs with comoving separations $<50$ Mpc to measure the expected signal from the mass distribution of LSS. However, in order to achieve the proper motion precisions in our catalog, astrometry was obtained for all quasars more than three times over at least a 10 year timespan. Therefore, addition of new close-separation pairs would be a long-term project that would require significant VLBI use. 

The Next Generation Very Large Array (ngVLA) provides another avenue for measuring new extragalactic proper motions. The ngVLA will provide ten times the effective collecting area and ten times longer baselines (300 km) than the current Karl Jansky Very Large Array \citep[JVLA;][]{Carillietal2015}. The larger collecting area will enable faster astrometry of fainter sources than is currently possible with the VLBA. In one hour of observation time, the ngVLA will detect objects with flux densities of a few tenths of a mJy or brighter \citep{Butleretal2018}.

If additional ngVLA antennae are installed at VLBA sites, the collecting power of the ngVLA could be combined with the astrometric precision of the VLBA to quickly create a large sample of close-separation pairs. For example, if 10,000 objects were monitored on a yearly basis to obtain $10 \ \mu$as yr$^{-1}$ astrometry per object, this would enable global detection of $0.1 \ \mu$as yr$^{-1}$ correlated signals at 5$\sigma$ significance \citep{Boweretal2015}. At this precision, we could constrain isotropy to 0.1\% of $H_0$ \citep{Boweretal2015}. If objects are selected to be in close-separation pairs, a new proper motion catalog with the ngVLA and VLBA would contain more than enough objects to detect the mass distribution of LSS; we predict that we need only $\sim 30$ pairs with separations $<50$ Mpc for a $3\sigma$ detection of this effect (Section \ref{future_LSS}).

However, it is uncertain whether the ngVLA will include long baselines $>1000$ km. For observations at 3.6 cm with a maximum baseline of 1000 km, the ngVLA will achieve a spatial resolution of $\sim 7$ mas. If objects are observed yearly for 10 years, then we expect to detect proper motions $> 700 \ \mu$as yr$^{-1}$. Therefore, unless the ngVLA includes baselines on order of VLBA baselines or can be used in conjunction with existing VLBI networks, the ngVLA will not be an effective tool to expand our sample of close-separation pairs. 

With the limitations of current VLBA catalogs and ngVLA baselines, {\it Gaia} proper motions remain the best avenue for increasing our sample of close-separation pairs. As discussed in Section \ref{future_LSS}, {\it Gaia} will produce $\sim 10^6$ new extragalactic proper motions \citep{Robinetal2012} with astrometric precisions of $\sim 1.69$ mas \citep{Lindegrenetal2016}. Although {\it Gaia} proper motions are much less precise \citep[uncertainties $\sim 200 \ \mu$as yr$^{-1}$;][]{Paineetal2017} than those produced through VLBI observations \citep[tens of $\mu$as yr$^{-1}$;][]{TitovLambert2013, TruebenbachDarling2017}, the large number of additional proper motions will allow a statistical detection of correlated signals. Additionally, optical proper motions typically have a smaller contribution from intrinsic proper motions than those measured in the radio because the optical source more frequently traces the galaxy core, rather than tracing a jet driven by the active galactic nucleus \citep{Darlingetal2017}.

In Section \ref{future_LSS}, we examined a subset of the {\it Gaia} DR1 catalog with 610,841 simulated extragalactic proper motions to estimate how well {\it Gaia} end-of-mission proper motions will be able to detect the relative velocity of close-separation pairs caused by the mass distribution of LSS. We found that the subset is able to significantly detect the predicted correlation of close-separation pairs. Therefore, although our current sample of close-separation pairs is too small to statistically detect the net convergence of gravitationally-interacting pairs, {\it Gaia} proper motions will be able to make a highly significant detection of the mass distribution of LSS on small scales. Comparison of the measured peculiar transverse two-point correlation statistic to that predicted by the theoretical matter power spectrum will allow a means to test the cosmological principles that contribute to the shape and magnitude of the power spectrum without a reliance on precise distance measurements.

\section*{Acknowledgments}
The authors thank David Gordon (NASA Goddard Space Flight Center), Nils Halverson, David Brain, Julie Comerford, and Alysia Marino (all University of Colorado Boulder) for helpful discussions. The authors acknowledge support from the NSF grant AST-1411605 and the NASA grant 14-ATP14-0086. 
All catalogue comparisons were performed using the tabular manipulation tool
STILTS \citep{Taylor2006}. This research has made use of NASA Goddard Space Flight
Center's VLBI source time series 2017a solution, prepared by
David Gordon. This research has made use of the NASA/IPAC Extragalactic Database (NED) which is operated by the Jet Propulsion Laboratory, California Institute of Technology, under contract with the National Aeronautics and Space Administration.

\software{LMFIT \citep{Newvilleetal2014}, STILTS \citep{Taylor2006}}

\newpage
\appendix

\section{Transverse Peculiar Velocity Two-Point Correlation Statistic}\label{appendix}

 For two objects in flat space at locations ($\alpha_1$,$\delta_1$,$x_1$) and ($\alpha_2$,$\delta_2$,$x_2$) in spherical coordinates, the vector connecting the two objects, converted to Cartesian coordinates, is
\begin{equation}
\vec{x} = 
\begin{pmatrix}
D_1 \cos \alpha_1 \cos \delta_1 - D_2 \cos \alpha_2 \cos \delta_2 \\
D_1 \sin \alpha_1 \cos \delta_1 - D_2 \sin \alpha_2 \cos \delta_2 \\
D_1 \sin \delta_1 - D_2 \sin \delta_2 
\end{pmatrix},
\end{equation}
where $D_1$ and $D_2$ are the proper distances of the two objects.

The transverse velocity of each object, in equatorial coordinates, is 
\begin{equation}
\vec{v}_{\bot} = D \ ( \mu_\alpha \ \hat{e}_\alpha + \mu_\delta \ \hat{e}_\delta ),
\end{equation}
where $\hat{e}_\alpha$ and $\hat{e}_\delta$ are the unit vectors along the right ascension and declination axes and $\mu_\alpha$ and $\mu_\delta$ are the object's proper motion in right ascension and declination. Using the conversion of $\hat{e}_\alpha$ and $\hat{e}_\delta$  to Cartesian coordinates from \citet{Mignard2012} (Eqns 5 and 6), 
\begin{equation}
\vec{v}_{\bot} = D
\begin{pmatrix}
-\mu_\alpha \sin \alpha - \mu_\delta \cos \alpha \sin \delta \\
\mu_\alpha \cos \alpha - \mu_\delta \sin \alpha \sin \delta \\
\mu_\delta \cos \delta
\end{pmatrix},
\end{equation}
with associated uncertainty
\begin{equation}
\sigma_{\vec{v}_{\bot}}^2 = D^2
\begin{pmatrix}
\sigma_{\mu_\alpha}^2 \sin^2 \alpha + \sigma_{\mu_\delta}^2 \cos^2 \alpha \sin^2 \delta \\
\sigma_{\mu_\alpha}^2 \cos^2 \alpha + \sigma_{\mu_\delta}^2 \sin^2 \alpha \sin^2 \delta \\
\sigma_{\mu_\delta}^2 \cos^2 \delta
\end{pmatrix}.
\end{equation}
We assume that the proper motion uncertainties are significantly larger than those in $\vec{x}$, $\alpha$, $\delta$, and $D$. 
\\

For each object, we compute $\vec{v}_{\bot} \cdot \hat{x}$ through a normal Cartesian dot product --
\begin{equation}
\vec{v}_{\bot} \cdot \hat{x} = \frac{1}{\lvert \vec{x} \rvert} \left ( v_{\bot,x} \ x_x + v_{\bot,y} \ x_y + v_{\bot,z} \ x_z  \right ).
\end{equation}
The uncertainty for the dot product is 
\begin{equation}
\sigma^2_{\vec{v}_{\bot} \cdot \hat{x}} =   \frac{1}{\lvert \vec{x} \rvert^2} \left (\sigma^2_{v_{\bot,x}} \ x^2_x + \sigma^2_{v_{\bot,y}} \ x^2_y + \sigma^2_{v_{\bot,z}} \ x^2_z \right ),
\end{equation} 
assuming negligible uncertainty in $\vec{x}$. Therefore, the correlation statistic for two individual objects is
\begin{equation}\label{xi_i}
\xi_i = (\vec{v}_{\bot,1} \cdot \hat{x})  (\vec{v}_{\bot,2} \cdot \hat{x}),
\end{equation}
with associated uncertainty
\begin{equation}
\sigma_{\xi_i}^2 = \xi_i^2 \left ( \frac{\sigma^2_{\vec{v}_{\bot,1} \cdot \hat{x}}}{(\vec{v}_{\bot,1} \cdot \hat{x})^2} + \frac{\sigma^2_{\vec{v}_{\bot,2} \cdot \hat{x}}}{(\vec{v}_{\bot,2} \cdot \hat{x})^2} \right ).
\end{equation}




\begin{thebibliography}{}

\bibitem[{{Bardeen} {et~al.}(1986){Bardeen}, {Bond}, {Kaiser}, \&
	{Szalay}}]{Bardeenetal1986}
{Bardeen}, J.~M., {Bond}, J.~R., {Kaiser}, N., \& {Szalay}, A.~S. 1986, \apj,
304, 15

\bibitem[{{Bastian}(1995)}]{Bastian1995}
{Bastian}, U. 1995, in ESA Special Publication, Vol. 379, Future Possibilities
for bstrometry in Space, ed. M.~A.~C. {Perryman} \& F.~{van Leeuwen}, 99

\bibitem[{{Blakeslee} {et~al.}(1999){Blakeslee}, {Ajhar}, \&
	{Tonry}}]{Blakesleeetal1999}
{Blakeslee}, J.~P., {Ajhar}, E.~A., \& {Tonry}, J.~L. 1999, in Astrophysics and
Space Science Library, Vol. 237, Post-Hipparcos Cosmic Candles, ed. A.~{Heck}
\& F.~{Caputo}, 181

\bibitem[{{Blumenthal} {et~al.}(1984){Blumenthal}, {Faber}, {Primack}, \&
	{Rees}}]{Blumenthaletal1984}
{Blumenthal}, G.~R., {Faber}, S.~M., {Primack}, J.~R., \& {Rees}, M.~J. 1984,
\nat, 311, 517

\bibitem[{{Bower} {et~al.}(2015){Bower}, {Demorest}, {Braatz}, {Broderick},
	{Burke-Spolaor}, {Butler}, {Chang}, {Chomiuk}, {Cordes}, {Darling}, {Eilek},
	{Hallinan}, {Kanekar}, {Kramer}, {Marrone}, {Max-Moerbeck}, {Metzger},
	{Morales}, {Myers}, {Osten}, {Owen}, {Rupen}, \& {Siemion}}]{Boweretal2015}
{Bower}, G.~C., {Demorest}, P., {Braatz}, J., {et~al.} 2015, ArXiv e-prints.


\bibitem[{{Bridle} \& {Perley}(1984)}]{BridlePerley1984}
{Bridle}, A.~H., \& {Perley}, R.~A. 1984, \araa, 22, 319

\bibitem[{{Butler} {et~al.}(2018){Butler}, {Grammer}, {Selina}, {Murphy}, \&
	{Carilli}}]{Butleretal2018}
{Butler}, B., {Grammer}, W., {Selina}, R., {Murphy}, E.~J., \& {Carilli}, C.
2018, in American Astronomical Society Meeting Abstracts, Vol. 231, American
Astronomical Society Meeting Abstracts, 342.09

\bibitem[{{Carilli} {et~al.}(2015){Carilli}, {McKinnon}, {Ott}, {Beasley},
	{Isella}, {Murphy}, {Leroy}, {Casey}, {Moullet}, {Lacy}, {Hodge}, {Bower},
	{Demorest}, {Hull}, {Hughes}, {di Francesco}, {Narayanan}, {Kent}, {Clark},
	\& {Butler}}]{Carillietal2015}
{Carilli}, C.~L., {McKinnon}, M., {Ott}, J., {et~al.} 2015, ArXiv e-prints.

\bibitem[{{Coles}(1993)}]{Coles1993}
{Coles}, P. 1993, \mnras, 262, 1065

\bibitem[{{Colgate}(1979)}]{Colgate1979}
{Colgate}, S.~A. 1979, \apj, 232, 404

\bibitem[{{Courtois} {et~al.}(2012){Courtois}, {Hoffman}, {Tully}, \&
	{Gottl{\"o}ber}}]{Courtoisetal2012}
{Courtois}, H.~M., {Hoffman}, Y., {Tully}, R.~B., \& {Gottl{\"o}ber}, S. 2012,
\apj, 744, 43

\bibitem[{{Darling}(2013)}]{Darling2013}
{Darling}, J. 2013, \apjl, 777, L21

\bibitem[{{Darling} \& {Truebenbach}(2018 submitted)}]{Darling2018}
{Darling}, J., \& {Truebenbach}, A.~E. 2018 submitted, {Connecting Proper
	Motions to the Matter Power Spectrum: Theory}

\bibitem[{{Darling} {et~al.}(2018 submitted){Darling}, {Truebenbach}, \&
	{Paine}}]{Darlingetal2017}
{Darling}, J., {Truebenbach}, A.~E., \& {Paine}, J. 2018 submitted, \apj

\bibitem[{{de Lapparent} {et~al.}(1986){de Lapparent}, {Geller}, \&
	{Huchra}}]{deLapparentetal1986}
{de Lapparent}, V., {Geller}, M.~J., \& {Huchra}, J.~P. 1986, \apjl, 302, L1

\bibitem[{{Dekel} {et~al.}(1993){Dekel}, {Bertschinger}, {Yahil}, {Strauss},
	{Davis}, \& {Huchra}}]{Dekeletal1993}
{Dekel}, A., {Bertschinger}, E., {Yahil}, A., {et~al.} 1993, \apj, 412, 1

\bibitem[{{Doumler} {et~al.}(2013){Doumler}, {Courtois}, {Gottl{\"o}ber}, \&
	{Hoffman}}]{Doumleretal2013}
{Doumler}, T., {Courtois}, H., {Gottl{\"o}ber}, S., \& {Hoffman}, Y. 2013

\bibitem[{{Eisenstein} \& {Hu}(1999)}]{EisensteinHu1999}
{Eisenstein}, D.~J., \& {Hu}, W. 1999, \apj, 511, 5

\bibitem[{{Eisenstein} {et~al.}(2005){Eisenstein}, {Zehavi}, {Hogg},
	{Scoccimarro}, {Blanton}, {Nichol}, {Scranton}, {Seo}, {Tegmark}, {Zheng},
	{Anderson}, {Annis}, {Bahcall}, {Brinkmann}, {Burles}, {Castander},
	{Connolly}, {Csabai}, {Doi}, {Fukugita}, {Frieman}, {Glazebrook}, {Gunn},
	{Hendry}, {Hennessy}, {Ivezi{\'c}}, {Kent}, {Knapp}, {Lin}, {Loh}, {Lupton},
	{Margon}, {McKay}, {Meiksin}, {Munn}, {Pope}, {Richmond}, {Schlegel},
	{Schneider}, {Shimasaku}, {Stoughton}, {Strauss}, {SubbaRao}, {Szalay},
	{Szapudi}, {Tucker}, {Yanny}, \& {York}}]{Eisensteinetal2005}
{Eisenstein}, D.~J., {Zehavi}, I., {Hogg}, D.~W., {et~al.} 2005, \apj, 633

\bibitem[{{Eubanks} {et~al.}(1995){Eubanks}, {Matsakis}, {Josties}, {Archinal},
	{Kingham}, {Martin}, {McCarthy}, {Klioner}, \& {Herring}}]{Eubanksetal1995}
{Eubanks}, T.~M., {Matsakis}, D.~N., {Josties}, F.~J., {et~al.} 1995, in IAU
Symposium, Vol. 166, Astronomical and Astrophysical Objectives of
Sub-Milliarcsecond Optical Astrometry, ed. E.~{Hog} \& P.~K. {Seidelmann},
283

\bibitem[{{Faber} \& {Jackson}(1976)}]{FaberJackson1976}
{Faber}, S.~M., \& {Jackson}, R.~E. 1976, \apj, 204, 668

\bibitem[{{Fanselow}(1983)}]{Fanselow1983}
{Fanselow}, J.~L. 1983, {Observation Model and parameter partial for the JPL
	VLBI parameter Estimation Software``MASTERFIT-V1.0''}, Tech. rep.

\bibitem[{{Fey} {et~al.}(1997){Fey}, {Eubanks}, \& {Kingham}}]{Feyetal1997}
{Fey}, A.~L., {Eubanks}, M., \& {Kingham}, K.~A. 1997, \aj, 114, 2284

\bibitem[{{Fry}(1996)}]{Fry1996}
{Fry}, J.~N. 1996, \apjl, 461, L65

\bibitem[{{Gaia Collaboration} {et~al.}(2018){Gaia Collaboration}, {Brown},
	{Vallenari}, {Prusti}, {de Bruijne}, {Babusiaux}, \&
	{Bailer-Jones}}]{GaiaCollaboration2018}
{Gaia Collaboration}, {Brown}, A.~G.~A., {Vallenari}, A., {et~al.} 2018, ArXiv
e-prints.


\bibitem[{{Gaia Collaboration} {et~al.}(2016){Gaia Collaboration}, {Prusti},
	{de Bruijne}, {Brown}, {Vallenari}, {Babusiaux}, {Bailer-Jones}, {Bastian},
	{Biermann}, {Evans}, \& et~al.}]{GaiaCollaboration2016}
{Gaia Collaboration}, {Prusti}, T., {de Bruijne}, J.~H.~J., {et~al.} 2016,
\aap, 595, A1

\bibitem[{{Gott} {et~al.}(2005){Gott}, {Juri{\'c}}, {Schlegel}, {Hoyle},
	{Vogeley}, {Tegmark}, {Bahcall}, \& {Brinkmann}}]{Gottetal2005}
{Gott}, III, J.~R., {Juri{\'c}}, M., {Schlegel}, D., {et~al.} 2005, \apj, 624,
463

\bibitem[{{Gwinn} {et~al.}(1997){Gwinn}, {Eubanks}, {Pyne}, {Birkinshaw}, \&
	{Matsakis}}]{Gwinnetal1997}
{Gwinn}, C.~R., {Eubanks}, T.~M., {Pyne}, T., {Birkinshaw}, M., \& {Matsakis},
D.~N. 1997, \apj, 485, 87

\bibitem[{{Hanes}(1977)}]{Hanes1977}
{Hanes}, D.~A. 1977, \memras, 84, 45

\bibitem[{{Harris} \& {Racine}(1979)}]{HarrisRacine1979}
{Harris}, W.~E., \& {Racine}, R. 1979, \araa, 17, 241

\bibitem[{{Hu} {et~al.}(1997){Hu}, {Spergel}, \& {White}}]{Huetal1997}
{Hu}, W., {Spergel}, D.~N., \& {White}, M. 1997, \prd, 55, 3288

\bibitem[{{Hu} \& {White}(1996{\natexlab{a}})}]{HuWhite1996a}
{Hu}, W., \& {White}, M. 1996{\natexlab{a}}, Physical Review Letters, 77, 1687

\bibitem[{{Hu} \& {White}(1996{\natexlab{b}})}]{HuWhite1996b}

\bibitem[{{Jensen} {et~al.}(1998){Jensen}, {Tonry}, \&
	{Luppino}}]{Jensenetal1998}
{Jensen}, J.~B., {Tonry}, J.~L., \& {Luppino}, G.~A. 1998, \apj, 505, 111

\bibitem[{{Kaiser}(1984)}]{Kaiser1984}
{Kaiser}, N. 1984, \apjl, 284, L9

\bibitem[{{Kitaura} {et~al.}(2012){Kitaura}, {Erdo{\v g}du}, {Nuza},
	{Khalatyan}, {Angulo}, {Hoffman}, \& {Gottl{\"o}ber}}]{Kitauraetal2012}
{Kitaura}, F.-S., {Erdo{\v g}du}, P., {Nuza}, S.~E., {et~al.} 2012, \mnras,
427, L35

\bibitem[{{Kopeikin} \& {Makarov}(2006)}]{KopeikinMakarov2006}
{Kopeikin}, S.~M., \& {Makarov}, V.~V. 2006, \aj, 131, 1471

\bibitem[{{Kovalevsky}(2003)}]{Kovalevsky2003}
{Kovalevsky}, J. 2003, \aap, 404, 743

\bibitem[{{Lindegren} {et~al.}(2016){Lindegren}, {Lammers}, {Bastian},
	{Hern{\'a}ndez}, {Klioner}, {Hobbs}, {Bombrun}, {Michalik}, {Ramos-Lerate},
	{Butkevich}, {Comoretto}, {Joliet}, {Holl}, {Hutton}, {Parsons},
	{Steidelm{\"u}ller}, {Abbas}, {Altmann}, {Andrei}, {Anton}, {Bach},
	{Barache}, {Becciani}, {Berthier}, {Bianchi}, {Biermann}, {Bouquillon},
	{Bourda}, {Br{\"u}semeister}, {Bucciarelli}, {Busonero}, {Carlucci},
	{Casta{\~n}eda}, {Charlot}, {Clotet}, {Crosta}, {Davidson}, {de Felice},
	{Drimmel}, {Fabricius}, {Fienga}, {Figueras}, {Fraile}, {Gai}, {Garralda},
	{Geyer}, {Gonz{\'a}lez-Vidal}, {Guerra}, {Hambly}, {Hauser}, {Jordan},
	{Lattanzi}, {Lenhardt}, {Liao}, {L{\"o}ffler}, {McMillan}, {Mignard}, {Mora},
	{Morbidelli}, {Portell}, {Riva}, {Sarasso}, {Serraller}, {Siddiqui}, {Smart},
	{Spagna}, {Stampa}, {Steele}, {Taris}, {Torra}, {van Reeven}, {Vecchiato},
	{Zschocke}, {de Bruijne}, {Gracia}, {Raison}, {Lister}, {Marchant},
	{Messineo}, {Soffel}, {Osorio}, {de Torres}, \&
	{O'Mullane}}]{Lindegrenetal2016}
{Lindegren}, L., {Lammers}, U., {Bastian}, U., {et~al.} 2016, \aap, 595, A4

\bibitem[{{Mei} {et~al.}(2005){Mei}, {Blakeslee}, {Tonry}, {Jord{\'a}n},
	{Peng}, {C{\^o}t{\'e}}, {Ferrarese}, {Merritt}, {Milosavljevi{\'c}}, \&
	{West}}]{Meietal2005}
{Mei}, S., {Blakeslee}, J.~P., {Tonry}, J.~L., {et~al.} 2005, \apjs, 156, 113

\bibitem[{{Mignard}(2002)}]{Mignard2002}
{Mignard}, F. 2002, in EAS Publications Series, Vol.~2, EAS Publications
Series, ed. O.~{Bienayme} \& C.~{Turon}, 327--339

\bibitem[{{Mignard} \& {Klioner}(2012)}]{Mignard2012}
{Mignard}, F., \& {Klioner}, S. 2012, \aap, 547, A59

\bibitem[{{Moodley} {et~al.}(2004){Moodley}, {Bucher}, {Dunkley}, {Ferreira},
	\& {Skordis}}]{Moodleyetal2004}
{Moodley}, K., {Bucher}, M., {Dunkley}, J., {Ferreira}, P.~G., \& {Skordis}, C.
2004, \prd, 70, 103520

\bibitem[{Newville {et~al.}(2014)Newville, Stensitzki, Allen, \&
	Ingargiola}]{Newvilleetal2014}
Newville, M., Stensitzki, T., Allen, D.~B., \& Ingargiola, A. 2014, {LMFIT:
	Non-Linear Least-Square Minimization and Curve-Fitting for Python}

\bibitem[{{Nusser} {et~al.}(2012){Nusser}, {Branchini}, \&
	{Davis}}]{Nusseretal2012}
{Nusser}, A., {Branchini}, E., \& {Davis}, M. 2012, \apj, 755, 58

\bibitem[{{Paine} {et~al.}(2018 submitted){Paine}, {Darling}, \&
	{Truebenbach}}]{Paineetal2017}
{Paine}, J., {Darling}, J., \& {Truebenbach}, A.~E. 2018 submitted, \apjs

\bibitem[{{Peiris} {et~al.}(2003){Peiris}, {Komatsu}, {Verde}, {Spergel},
	{Bennett}, {Halpern}, {Hinshaw}, {Jarosik}, {Kogut}, {Limon}, {Meyer},
	{Page}, {Tucker}, {Wollack}, \& {Wright}}]{Peirisetal2003}
{Peiris}, H.~V., {Komatsu}, E., {Verde}, L., {et~al.} 2003, \apjs, 148, 213

\bibitem[{{Perlmutter} {et~al.}(1999){Perlmutter}, {Aldering}, {Goldhaber},
	{Knop}, {Nugent}, {Castro}, {Deustua}, {Fabbro}, {Goobar}, {Groom}, {Hook},
	{Kim}, {Kim}, {Lee}, {Nunes}, {Pain}, {Pennypacker}, {Quimby}, {Lidman},
	{Ellis}, {Irwin}, {McMahon}, {Ruiz-Lapuente}, {Walton}, {Schaefer}, {Boyle},
	{Filippenko}, {Matheson}, {Fruchter}, {Panagia}, {Newberg}, {Couch}, \&
	{Project}}]{Perlmutteretal1999}
{Perlmutter}, S., {Aldering}, G., {Goldhaber}, G., {et~al.} 1999, \apj, 517

\bibitem[{{Planck Collaboration} {et~al.}(2016{\natexlab{a}}){Planck
		Collaboration}, {Adam}, {Ade}, {Aghanim}, {Akrami}, {Alves}, {Arg{\"u}eso},
	{Arnaud}, {Arroja}, {Ashdown}, \& et~al.}]{PlanckI2016}
{Planck Collaboration}, {Adam}, R., {Ade}, P.~A.~R., {et~al.}
2016{\natexlab{a}}, \aap, 594, A1

\bibitem[{{Planck Collaboration} {et~al.}(2016{\natexlab{b}}){Planck
		Collaboration}, {Ade}, {Aghanim}, {Arnaud}, {Ashdown}, {Aumont},
	{Baccigalupi}, {Banday}, {Barreiro}, {Bartlett}, \& et~al.}]{PlanckXIII2016}
{Planck Collaboration}, {Ade}, P.~A.~R., {Aghanim}, N., {et~al.}
2016{\natexlab{b}}, \aap, 594, A13

\bibitem[{{Planck Collaboration} {et~al.}(2016{\natexlab{c}}){Planck
		Collaboration}, {Ade}, {Aghanim}, {Arnaud}, {Arroja}, {Ashdown}, {Aumont},
	{Baccigalupi}, {Ballardini}, {Banday}, \& et~al.}]{PlanckXX2016}
---. 2016{\natexlab{c}}, \aap, 594, A20

\bibitem[{{Planck Collaboration} {et~al.}(2016{\natexlab{d}}){Planck
		Collaboration}, {Ade}, {Aghanim}, {Arnaud}, {Ashdown}, {Aumont},
	{Baccigalupi}, {Banday}, {Barreiro}, {Bartolo}, \& et~al.}]{PlanckXIV2016}
---. 2016{\natexlab{d}}, \aap, 594, A14

\bibitem[{{Popovi{\'c}} {et~al.}(2012){Popovi{\'c}}, {Jovanovi{\'c}},
	{Stalevski}, {Anton}, {Andrei}, {Kova{\v c}evi{\'c}}, \&
	{Baes}}]{Popovicetal2012}
{Popovi{\'c}}, L.~{\v C}., {Jovanovi{\'c}}, P., {Stalevski}, M., {et~al.} 2012,
\aap, 538, A107

\bibitem[{{Popovi{\'c}} \& {Simi{\'c}}(2013)}]{PopovicSimic2013}
{Popovi{\'c}}, L.~{\v C}., \& {Simi{\'c}}, S. 2013, \mnras, 432, 848

\bibitem[{{Quercellini} {et~al.}(2009){Quercellini}, {Quartin}, \&
	{Amendola}}]{Quercellinietal2009}
{Quercellini}, C., {Quartin}, M., \& {Amendola}, L. 2009, Physical Review
Letters, 102, 151302

\bibitem[{{Reid} {et~al.}(2009){Reid}, {Menten}, {Zheng}, {Brunthaler},
	{Moscadelli}, {Xu}, {Zhang}, {Sato}, {Honma}, {Hirota}, {Hachisuka}, {Choi},
	{Moellenbrock}, \& {Bartkiewicz}}]{Reidetal2009}
{Reid}, M.~J., {Menten}, K.~M., {Zheng}, X.~W., {et~al.} 2009, \apj, 700, 137

\bibitem[{{Riess} {et~al.}(1996){Riess}, {Press}, \&
	{Kirshner}}]{Riessetal1996}
{Riess}, A.~G., {Press}, W.~H., \& {Kirshner}, R.~P. 1996, \apj, 473, 88

\bibitem[{{Riess} {et~al.}(2001){Riess}, {Nugent}, {Gilliland}, {Schmidt},
	{Tonry}, {Dickinson}, {Thompson}, {Budav{\'a}ri}, {Casertano}, {Evans},
	{Filippenko}, {Livio}, {Sanders}, {Shapley}, {Spinrad}, {Steidel}, {Stern},
	{Surace}, \& {Veilleux}}]{Riessetal2001}
{Riess}, A.~G., {Nugent}, P.~E., {Gilliland}, R.~L., {et~al.} 2001, \apj, 560,
49

\bibitem[{{Riess} {et~al.}(2011){Riess}, {Macri}, {Casertano}, {Lampeitl},
	{Ferguson}, {Filippenko}, {Jha}, {Li}, \& {Chornock}}]{Riessetal2011}
{Riess}, A.~G., {Macri}, L., {Casertano}, S., {et~al.} 2011, \apj, 730, 119

\bibitem[{{Riess} {et~al.}(2016){Riess}, {Macri}, {Hoffmann}, {Scolnic},
	{Casertano}, {Filippenko}, {Tucker}, {Reid}, {Jones}, {Silverman},
	{Chornock}, {Challis}, {Yuan}, {Brown}, \& {Foley}}]{Riessetal2016}
{Riess}, A.~G., {Macri}, L.~M., {Hoffmann}, S.~L., {et~al.} 2016, \apj, 826,
56

\bibitem[{{Riess} {et~al.}(2018){Riess}, {Casertano}, {Yuan}, {Macri},
	{Bucciarelli}, {Lattanzi}, {MacKenty}, {Bowers}, {Zheng}, {Filippenko},
	{Huang}, \& {Anderson}}]{Riessetal2018}
{Riess}, A.~G., {Casertano}, S., {Yuan}, W., {et~al.} 2018, ArXiv e-prints.

\bibitem[{{Robin} {et~al.}(2012){Robin}, {Luri}, {Reyl{\'e}}, {Isasi}, {Grux},
	{Blanco-Cuaresma}, {Arenou}, {Babusiaux}, {Belcheva}, {Drimmel}, {Jordi},
	{Krone-Martins}, {Masana}, {Mauduit}, {Mignard}, {Mowlavi},
	{Rocca-Volmerange}, {Sartoretti}, {Slezak}, \& {Sozzetti}}]{Robinetal2012}
{Robin}, A.~C., {Luri}, X., {Reyl{\'e}}, C., {et~al.} 2012, \aap, 543, A100

\bibitem[{Sivia \& Skilling(2006)}]{SiviaSkilling2006}
Sivia, D., \& Skilling, J. 2006, Data analysis: a Bayesian tutorial, Oxford
science publications (Oxford University Press).
\bibitem[{{Souchay} {et~al.}(2009){Souchay}, {Andrei}, {Barache}, {Bouquillon},
	{Gontier}, {Lambert}, {Le Poncin-Lafitte}, {Taris}, {Arias}, {Suchet}, \&
	{Baudin}}]{Souchayetal2009}
{Souchay}, J., {Andrei}, A.~H., {Barache}, C., {et~al.} 2009, \aap, 494, 799

\bibitem[{{Souchay} {et~al.}(2015){Souchay}, {Andrei}, {Barache}, {Kalewicz},
	{Gattano}, {Coelho}, {Taris}, {Bouquillon}, \& {Becker}}]{Souchayetal2015}
---. 2015, \aap, 583, A75

\bibitem[{{Sovers} {et~al.}(1998){Sovers}, {Fanselow}, \&
	{Jacobs}}]{Soversetal1998}
{Sovers}, O.~J., {Fanselow}, J.~L., \& {Jacobs}, C.~S. 1998, Reviews of Modern
Physics, 70, 1393

\bibitem[{{Strauss} {et~al.}(1992){Strauss}, {Davis}, {Yahil}, \&
	{Huchra}}]{Straussetal1992}
{Strauss}, M.~A., {Davis}, M., {Yahil}, A., \& {Huchra}, J.~P. 1992, \apj, 385,
421

\bibitem[{{Taylor}(2006)}]{Taylor2006}
{Taylor}, M.~B. 2006, in Astronomical Society of the Pacific Conference Series,
Vol. 351, Astronomical Data Analysis Software and Systems XV, ed.
C.~{Gabriel}, C.~{Arviset}, D.~{Ponz}, \& S.~{Enrique}, 666

\bibitem[{{Tegmark} \& {Peebles}(1998)}]{TegmarkPeebles1998}
{Tegmark}, M., \& {Peebles}, P.~J.~E. 1998, \apjl, 500, L79

\bibitem[{{Titov} \& {Lambert}(2013)}]{TitovLambert2013}
{Titov}, O., \& {Lambert}, S. 2013, \aap, 559, A95

\bibitem[{{Titov} {et~al.}(2011){Titov}, {Lambert}, \&
	{Gontier}}]{Titovetal2011}
{Titov}, O., {Lambert}, S.~B., \& {Gontier}, A.-M. 2011, \aap, 529, A91

\bibitem[{{Tonry} \& {Schneider}(1988)}]{TonrySchneider1988}
{Tonry}, J., \& {Schneider}, D.~P. 1988, \aj, 96, 807

\bibitem[{{Tonry} {et~al.}(1990){Tonry}, {Ajhar}, \& {Luppino}}]{Tonryetal1990}
{Tonry}, J.~L., {Ajhar}, E.~A., \& {Luppino}, G.~A. 1990, \aj, 100, 1416

\bibitem[{{Treyer} \& {Wambsganss}(2004)}]{TreyerWambsganss2004}
{Treyer}, M., \& {Wambsganss}, J. 2004, \aap, 416, 19

\bibitem[{{Truebenbach} \& {Darling}(2017)}]{TruebenbachDarling2017}
{Truebenbach}, A.~E., \& {Darling}, J. 2017, \apjs, 233, 3

\bibitem[{{Tully} {et~al.}(2014){Tully}, {Courtois}, {Hoffman}, \&
	{Pomar{\`e}de}}]{Tullyetal2014}
{Tully}, R.~B., {Courtois}, H., {Hoffman}, Y., \& {Pomar{\`e}de}, D. 2014,
\nat, 513, 71

\bibitem[{{Xu} {et~al.}(2012){Xu}, {Wang}, \& {Zhao}}]{Xuetal2012}
{Xu}, M.~H., {Wang}, G.~L., \& {Zhao}, M. 2012, \aap, 544, A135

\bibitem[{{Xu} {et~al.}(2013){Xu}, {Wang}, \& {Zhao}}]{Xuetal2013}
---. 2013, \mnras, 430, 2633

\bibitem[{{York} {et~al.}(2000){York}, {Adelman}, {Anderson}, {Anderson},
	{Annis}, {Bahcall}, {Bakken}, {Barkhouser}, {Bastian}, {Berman}, {Boroski},
	{Bracker}, {Briegel}, {Briggs}, {Brinkmann}, {Brunner}, {Burles}, {Carey},
	{Carr}, {Castander}, {Chen}, {Colestock}, {Connolly}, {Crocker}, {Csabai},
	{Czarapata}, {Davis}, {Doi}, {Dombeck}, {Eisenstein}, {Ellman}, {Elms},
	{Evans}, {Fan}, {Federwitz}, {Fiscelli}, {Friedman}, {Frieman}, {Fukugita},
	{Gillespie}, {Gunn}, {Gurbani}, {de Haas}, {Haldeman}, {Harris}, {Hayes},
	{Heckman}, {Hennessy}, {Hindsley}, {Holm}, {Holmgren}, {Huang}, {Hull},
	{Husby}, {Ichikawa}, {Ichikawa}, {Ivezi{\'c}}, {Kent}, {Kim}, {Kinney},
	{Klaene}, {Kleinman}, {Kleinman}, {Knapp}, {Korienek}, {Kron}, {Kunszt},
	{Lamb}, {Lee}, {Leger}, {Limmongkol}, {Lindenmeyer}, {Long}, {Loomis},
	{Loveday}, {Lucinio}, {Lupton}, {MacKinnon}, {Mannery}, {Mantsch}, {Margon},
	{McGehee}, {McKay}, {Meiksin}, {Merelli}, {Monet}, {Munn}, {Narayanan},
	{Nash}, {Neilsen}, {Neswold}, {Newberg}, {Nichol}, {Nicinski}, {Nonino},
	{Okada}, {Okamura}, {Ostriker}, {Owen}, {Pauls}, {Peoples}, {Peterson},
	{Petravick}, {Pier}, {Pope}, {Pordes}, {Prosapio}, {Rechenmacher}, {Quinn},
	{Richards}, {Richmond}, {Rivetta}, {Rockosi}, {Ruthmansdorfer}, {Sandford},
	{Schlegel}, {Schneider}, {Sekiguchi}, {Sergey}, {Shimasaku}, {Siegmund},
	{Smee}, {Smith}, {Snedden}, {Stone}, {Stoughton}, {Strauss}, {Stubbs},
	{SubbaRao}, {Szalay}, {Szapudi}, {Szokoly}, {Thakar}, {Tremonti}, {Tucker},
	{Uomoto}, {Vanden Berk}, {Vogeley}, {Waddell}, {Wang}, {Watanabe},
	{Weinberg}, {Yanny}, {Yasuda}, \& {SDSS Collaboration}}]{Yorketal2000}
{York}, D.~G., {Adelman}, J., {Anderson}, Jr., J.~E., {et~al.} 2000, \aj, 120,
1579

\bibitem[{{Zhang} {et~al.}(2017){Zhang}, {Childress}, {Davis}, {Karpenka},
	{Lidman}, {Schmidt}, \& {Smith}}]{Zhangetal2017}
{Zhang}, B.~R., {Childress}, M.~J., {Davis}, T.~M., {et~al.} 2017, \mnras, 471,
2254
	
\end{thebibliography}


\end{document}